\documentclass[conference,a4paper]{APSIPA2021}
\usepackage{amsmath}
\usepackage{graphicx}
\usepackage{multirow}
\usepackage{threeparttable}

\usepackage{chngcntr}

\newcounter{mysfig}
\counterwithin{mysfig}{figure}

\renewcommand\themysfig{(\alph{mysfig})}
\makeatletter
\newcommand\Scaption[1]{%
\refstepcounter{mysfig}%
\vskip.5\abovecaptionskip
  \sbox\@tempboxa{\small\themysfig~#1}%
  \ifdim \wd\@tempboxa >\hsize
    \small\themysfig~#1\par
  \else
    \global \@minipagefalse
    \hb@xt@\hsize{\hfil\box\@tempboxa\hfil}%
  \fi
  \vskip\belowcaptionskip}
\makeatother

\usepackage{cleveref}
\usepackage{amsfonts,amssymb,amsthm, bm}

\usepackage[style=ieee,]{biblatex}
\usepackage{geometry}
\usepackage{fancyhdr}

\fancypagestyle{firststyle}{
  \fancyhf{}
  \fancyhead[C]{2024 Asia Pacific Signal and Information Processing Association Annual Summit and Conference (APSIPA ASC)}
}

\begin{document}

\title{Music2Fail: Transfer Music to Failed Recorder Style}





\author{
\authorblockN{
Chon In Leong, I-Ling Chung, 
 Kin-Fong Chao, \\ Jun-You Wang, Yi-Hsuan Yang, Jyh-Shing Roger Jang
}

\authorblockA{
National Taiwan University, Taiwan \\
E-mail: r11922174@csie.ntu.edu.tw, r11942103@ntu.edu.tw, r12922179@ntu.edu.tw, \\ junyou.wang@mirlab.org, yhyangtw@ntu.edu.tw, jang@csie.ntu.edu.tw}
}

\maketitle
\thispagestyle{firststyle}
\pagestyle{fancy}

\pagestyle{empty}

\begin{abstract}
  The goal of music style transfer is to convert a music performance by one instrument into another while keeping the musical contents unchanged. In this paper, we investigate another style transfer scenario called ``failed-music style transfer''. Unlike the usual music style transfer where the content remains the same and only the instrumental characteristics are changed, this scenario seeks to transfer the music from the source instrument to the target instrument which is deliberately performed off-pitch. Our work attempts to transfer normally played music into off-pitch recorder music, which we call ``failed-style recorder'', and study the results of the conversion. 
  To carry out this work, we have also proposed a dataset of failed-style recorders for this task, called ``FR109 Dataset''.
  Such an experiment explores the music style transfer task in a more expressive setting, as the generated audio should sound like an ``off-pitch recorder'' while maintaining a certain degree of naturalness.~\footnote{Our demo website is now available on: \url{https://navi0105.github.io/demo/music2fail/}}
  
\end{abstract}

\section{Introduction}\label{sec:introduction}
    Generally, the goal of \emph{music style transfer} is to change the style of the input audio while preserving the content of the input audio\footnote{In practice, due to the differences in pitch ranges of different instruments, we may perform additional pitch shifting in the experiments.}. 
    In particular, the \emph{content} of the input music data refers to music features such as rhythm and melody, while the \emph{style} refers to the unique perceived characteristics that an instrument expresses in music performance.
    According to the wide range of conversion goals and targets, the tasks of music style transfer can be broadly classified into three categories: \emph{one-to-one}~\cite{pasini2019melgan, huang2018timbretron}, \emph{many-to-one}~\cite{engel2020ddsp, Nercessian2022differentiable}, \emph{many-to-many}~\cite{bitton2018modulated, wu2023transplayer}. Several music style transfer methods have been inspired by research from different fields, such as voice conversion (VC)~\cite{wu2023transplayer, bonnici2022timbre, bitton2018modulated, comanducci2023timbre} and image-based style transfer~\cite{kaneko2018cyclegan, liu2018unsupervised}.
    Most works mainly use well-played audio as training data for both the source and target domains. This implies that well-played input audio will be converted into equally well-played output audio. Although this ensures faithful style transfer, assuming all audio to be well-played in the real world restricts the expressiveness of instruments. The Singing Voice Beautifying (SVB) task~\cite{liu-etal-2022-learning-beauty} is a special case. They used paired data of amateur and professional singing voices for training, aiming to correct the pitch and improve the vocal tone, indicating that the source audio was not well-played. In this paper, we focus on another special case, where the target domain is not well-played. For example, the target domain may be a soprano recorder that is deliberately performed poorly\footnote{A famous example can be found at \url{https://www.youtube.com/watch?v=X2WH8mHJnhM}}. We refer to this case as \emph{failed-music style transfer}. The motivation is that such failed music contains a wider range of characteristics, but humans can still distinguish between a ``failed recorder'' and another instrument. This poses a more difficult scenario to music style transfer: 
    Can a music style transfer model not only tackle well-played instruments but also instruments that are \emph{not} well-played?

    Take a soprano recorder as an example, a fail-style recorder may contain many types of errors, such as:
    \begin{itemize}
    \itemsep=-1pt
        \item {\bf Cracked voice.} Producing a harsh sound.
        \item {\bf Weird dynamics.} Unnatural volume while playing.
        \item {\bf Failed tonguing.} Mistakes in the articulation.
        \item {\bf Overblowing.} Blowing too hard, causing the voice to sound raspy.
        \item {\bf Underblowing.} Blowing not hard enough, causing the voice to sound hissing.
    \end{itemize} 
    
    Generally, these are considered errors that should not occur in live performances. However, by definition, such errors should not make a style transfer model malfunction. Instead, a style transfer model should generate audio that sounds like a failed recorder (sometimes has an unpleasant style, but still sounds like a recorder). Such failed music style transfer might be useful in the fields of entertainment, it could serve as the score for some comedies or some humorous scenes.
    
In this paper, we investigate the music style transfer scenario of failed recorders, treating it as a type of instrument. We apply various general style transfer methods and analyze the conversion results. However, there are no existing datasets for such scenarios which were deliberately recorded as failed-style. To facilitate our research, we propose the ``FR109'' dataset, a collection of failed-style recorder music recorded by a professional, with deliberately included failures.

    To sum up, the main contributions of this paper are two-fold:
    \begin{itemize}
        \item 
        We discuss the special scenario of \emph{failed-music style transfer} that serves as a more challenging task for style transfer. Regarding the experimental results, we analyzed them from the perspectives of the Mel spectrogram and Wiener entropy, providing corresponding analyses and interpretations.
        \item 
        To carry out the work for \emph{failed-music style transfer}, we propose the FR109 dataset, a dataset of failed recorder performance, created intentionally by an experienced individual playing a recorder.
    \end{itemize}

\section{Datasets}\label{sec:datasets}
    In this work, we adopted three datasets in the experiments, including two publicly available datasets (URMP~\cite{li2018creating} and Bach10~\cite{Duan2010multiple}),
    and our FR109 dataset.
    The comparison of different datasets is shown in Table~\ref{tab:datasets}.

\subsection{The URMP dataset}\label{subsec:urmp}
    The URMP dataset \cite{li2018creating} contains 44 music pieces ranging from duets to quintets, with separated tracks for individual instrument recordings. There are 14 distinct instruments in this dataset. 
    In our experiments, we only used the violin, clarinet, and saxophone tracks as the training data.

\subsection{The Bach10 dataset}\label{subsec:bach10}
    The Bach10 dataset \cite{Duan2010multiple} consists of audio recordings of 10 J.S. Bach chorales performed separately with violin, clarinet, saxophone, and bassoon. In our experiments, we only used the violin, clarinet, and saxophone tracks as the testing data. 
    Since the training data (URMP) and testing data (Bach10) belong to different datasets, such an evaluation scenario is more challenging.

\subsection{The proposed FR109 dataset}\label{subsec:fr109}
    As for the failed-music style transfer, 
    we proposed the FR109 dataset, 
    which consists of 109 songs recorded with a soprano recorder played by a professional, with a total duration of 5.05 hours.
    Errors are introduced to each performance intentionally.
    
    As discussed in Section~\ref{sec:introduction}, the types of errors include cracked voice, weird dynamics, failed tonguing, overblowing, and underblowing. To compute the statistics of the dataset, we extract the pitch of recorder music using CREPE~\cite{kim2018crepe}. The pitch mean of the FR109 dataset is around 905Hz (between A5 and A\#5), and the maximum pitch value is 1990Hz (around B6). These statistics match the actual pitch range of the soprano recorder, which spans from C5 to D7.

    Since there is no other dataset for failed recorders, we use the FR109 dataset as both the training dataset and the testing dataset in the failed-music style transfer experiments. A 90\%/10\% split is employed to divide the dataset into a training dataset and a testing dataset. In our experiments, we trained style transfer models to perform style transfer between all 4 instruments (violin, clarinet, saxophone, and failed recorder).

    We are making the FR109 dataset publicly available for reproducibility, you can check out more information on our official GitHub page: \url{https://github.com/navi0105/Music2Fail} 

    \begin{table}[t]
        \begin{center}
        \begin{threeparttable}
            \caption{The list of datasets utilized in this work, including the proposed FR109 dataset. ``Pieces'' stands for the number of music pieces.}
            \label{tab:datasets}
            \begin{tabular}{|c|c|c|c|}
                \hline
                Dataset                 & Instrument & Pieces & Total duration \\ \hline\hline
                \multirow{3}{*}{URMP}   & Violin & 34 & 1.02 hours \\ \cline{2-4} 
                                        & Clarinet & 10 & 0.30 hours\\ \cline{2-4} 
                                        & Saxophone & 11 & 0.26 hours \\ \hline\hline
                \multirow{3}{*}{Bach10} & Violin & 10 & 0.09 hours \\ \cline{2-4} 
                                        & Clarinet & 10 & 0.09 hours \\ \cline{2-4} 
                                        & Saxophone & 10 & 0.09 hours \\ \hline\hline
                FR109 & (Failed) recorder & 109 & 5.05 hours \\ \hline
        \end{tabular}
        
        \end{threeparttable}
        \end{center}
    \end{table}

\section{Method}\label{sec:method}
    In this work, we experiment with three different well-known style transfer methods for failed-music style transfer, they are StarGAN \cite{choi2018stargan}, VAE-GAN \cite{bonnici2022timbre}, and DDSP \cite{engel2020ddsp}.

    StarGAN \cite{choi2018stargan} introduced domain labels to the generator and discriminator, 
    the generator uses the domain label to specify the target domain, while the discriminator needs to predict the input's domain. 
    During training, the generator and the discriminator contest with each other, 
    the generator's objective is to fool the discriminator, making it mispredict the domain label, while the discriminator's objective is to avoid being fooled by the generator.
    StarGAN only used a single generator and discriminator for learning a multi-mapping between different styles, instead of one generator and one discriminator for each pair of styles.

    VAE-GAN \cite{bonnici2022timbre} used one generator and one discriminator for each domain, the generator used a variational autoencoder composed of two parts: the universal encoder and a decoder, the universal encoder shared across each generator to encode the input to latent code, and a decoder to transfer the latent code to the target domain. 
    Since it uses the same encoder for every input domain and target domain, the performance was increased due to the variation of the input data.
    The decoder is domain-specific so it can be specialized to that domain.
    The discriminator only needed to predict whether the data was generated for that domain.

    DDSP \cite{engel2020ddsp} integrates classic signal processing with deep learning. This method employs an autoencoder architecture for style transfer within a single domain. The encoder extracts key features from the source audio, including loudness, fundamental frequency, and residual information, while the decoder maps these features to control parameters for synthesizers to generate the output audio. DDSP has an assumption that the pitch component extracted from the source audio should closely match the fundamental frequency of the output audio, which may not be suitable for failed-music style transfer.

    StarGAN and VAE-GAN are two-stage style transfer pipelines, where we first convert the source audio to Mel spectrogram. Then, the Mel spectrogram was transferred to the failed recorder style using the generator. Finally, the vocoder generates waveform from the transferred Mel spectrogram. Here, we use BigVSAN \cite{shibuya2024bigvsan} as our vocoder, its pretrained weight are available in their official repository\footnote{https://github.com/sony/bigvsan}, which was pretrained on the LibriTTS dataset's training dataset~\cite{Zen2019libritts} for 10 million steps.

\section{Experiments}\label{sec:experiments}
    As discussed in Section~\ref{sec:datasets}, we combined the URMP dataset~\cite{li2018creating} and the FR109 dataset's training dataset for training and the Bach10 dataset~\cite{Duan2010multiple} for evaluation. Three different methods are compared in the experiments, they are StarGAN \cite{choi2018stargan}, VAE-GAN \cite{bonnici2022timbre}, and DDSP \cite{engel2020ddsp}. 


\subsection{Training}

\subsubsection{Data preprocessing}\label{subsec:data_pre}
    We refer to the arguments for calculating the Mel spectrogram from BigVSAN~\cite{shibuya2024bigvsan} to compute the Mel spectrograms of music data in our dataset, 24,000 for sampling rate, 100-bands of Mel filter bank, 1024 for FFT / Hann window, hop size is 256 and the frequency range is from 0 to 12,000 Hz.



\subsection{Evaluation}\label{subsec:evaluation}
    In this section, we compare the performance between StarGAN, VAE-GAN, and DDSP. Both objective and subjective experiments are conducted.

\subsubsection{Objective evaluation}
    Fr\'echet Audio Distance (FAD) \cite{kilgour2018fr} is a reference-free metric to compute the Fr\'echet Inception Distance (FID) between audio embedding sets extracted from the reference set and evaluation set. In the experiments, the reference set is music from the testing dataset and the evaluation set is music generated by the model. FAD represents the degree of dissimilarity between the two sets. The audio embeddings are extracted by a pretrained VGGish audio classification model~\cite{hershey2017cnn}.
    We use FAD as an objective evaluation metric to assess the distance between the audio files converted by StarGAN / VAE-GAN / DDSP and the real audio performance of a target instrument. 

    \Cref{tab:FAD_torecorder} shows the FAD score of Bach10 dataset's music converted to failed-style recorder music using these three different models.
    The results indicate that StarGAN performs slightly worse than VAE-GAN on both datasets. Considering that StarGAN only utilizes one unified decoder while VAE-GAN uses one decoder for each instrument, such a performance gap is acceptable.
    
    As for the DDSP model, results show that DDSP has a significantly larger FAD compared to StarGAN and VAR-GAN. By inspecting the audio converted by DDSP, we found that they contain a large amount of noise, which is the reason for the high FAD. 
    The results of DDSP demonstrate that its assumptions about pitch invariance can lead to better performance on well-played instrument transitions, but do not apply well to our task.


    \begin{table}[t]
    \setlength{\tabcolsep}{24pt}
        \begin{center}
        \begin{threeparttable}
        \caption{The FAD metrics of style transfer models to failed recorder in the Bach10 Dataset.}
        \label{tab:FAD_torecorder}
        \begin{tabular}{|c|c|}
            \hline
            Models & FAD ($\downarrow$) \\ 
            \hline
            \hline
            StarGAN & 13.87 \\ \hline
            VAE-GAN & \textbf{7.27} \\ \hline
            DDSP & 38.91 \\ \hline
    
        \end{tabular}
    \end{threeparttable}
    \end{center}
    \end{table}

\subsubsection{Subjective evaluation}\label{subsubsec:subjective_eval}
    We performed a listening test that evaluates the performance of converting these three (well-played) instruments into failed recorder in the Bach10 dataset (3 source-target pairs).
    
    For each source-target pair, we randomly choose one audio clip for the listening test. We employed a rating scheme based on the Mean Opinion Score (MOS) \cite{streijl2016mean}. For each audio clip (converted by one of the models), we asked the participants to evaluate its quality in three aspects: (1) \textit{Style similarity} ({SS}) to the target instrument, (2) \textit{Melody similarity} ({MS}) to the source audio, (3) \textit{Sound quality} ({SQ}) of the converted audio. The scoring ranges from 1 to 5, where 1 is the worst and 5 is the best. In total, we received 16 valid responses from the listening test. 
    \Cref{tab:MOS_b} shows the MOS of the Bach10 dataset.


    The results indicate that StarGAN's overall performance falls behind VAE-GAN's on all metrics in the conversion to failed recorder. The $p$-values between StarGAN and VAE-GAN are 0.09 (SS), 0.06 (MS), and 0.02 (SQ). Although only the sound quality (SQ) is considered statistically significant, overall, we can still conclude that StarGAN is slightly inferior to VAE-GAN in converting to failed recorder music.
    As for DDSP, similar to the objective results, the MOS results of DDSP are significantly worse than those of StarGAN. This is likely because DDSP generates noise more frequently. All the $t$-test yielded $p$-values well below 0.05. This reflects the specific challenges involved in music style transfer to failed instrument music for DSP-based synthesizers.
    

    \begin{table}[t]
    \begin{center}
    \begin{threeparttable}
    \setlength{\tabcolsep}{5pt}
        \caption{The MOS of the listening test on the Bach10 dataset. The numbers inside the cells represent the MOS and their standard deviations. SS, MS, and SQ indicate the style similarity, melody similarity, and sound quality, respectively.}
        \label{tab:MOS_b}
        \begin{tabular}{|c|c|c|c|}
            \hline
            Models & SS ($\uparrow$) & MS ($\uparrow$) & SQ ($\uparrow$) \\
            \hline
            \hline
            StarGAN & 2.54 $\pm$ 1.26 & 3.15 $\pm$ 1.19 & 2.46 $\pm$ 1.27 \\
            \hline
            VAE-GAN & \textbf{2.98} $\pm$ 1.23 & \textbf{3.56} $\pm$ 0.93 & \textbf{3.00} $\pm$ 0.98 \\
            \hline
            DDSP & 1.33 $\pm$ 0.77 & 2.19 $\pm$ 1.11 & 1.38 $\pm$ 0.70 \\
            \hline
        \end{tabular}
        \end{threeparttable}
    \end{center}
    \end{table}

\begin{figure}[t]
    \centering
    \includegraphics[width=0.15\textwidth, trim= 250 30 250 50, clip]{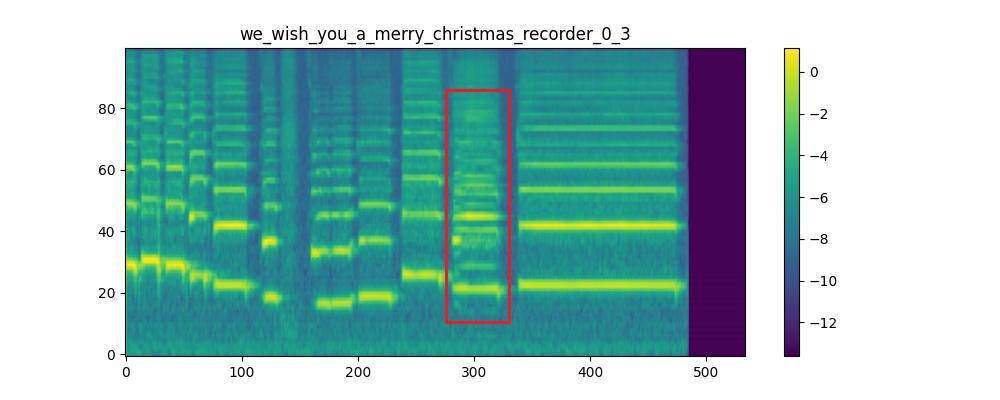}
    \includegraphics[width=0.15\textwidth, trim= 100 30 400 50, clip]{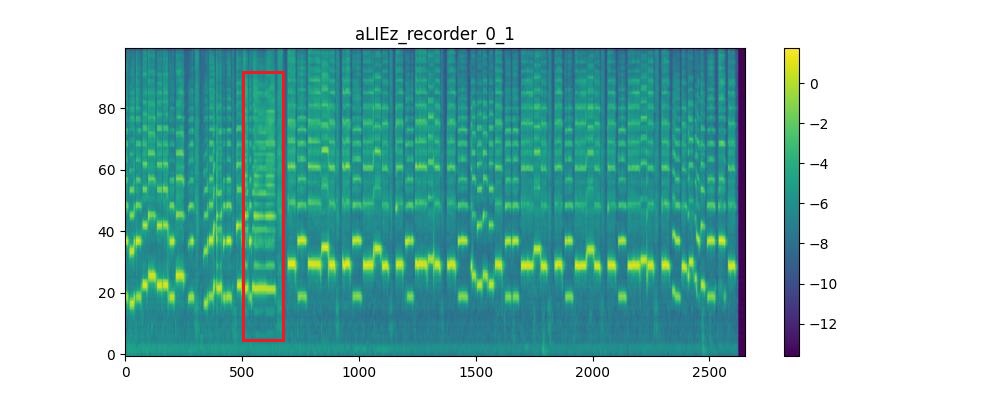}
    \includegraphics[width=0.08\textwidth, trim= 400 30 200 45, clip]{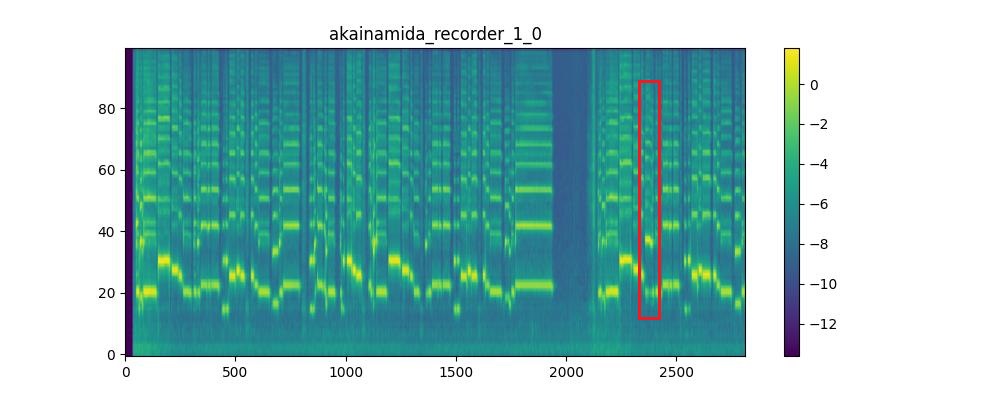}
    \caption{Mel spectrograms of failed recorder music, the red rectangular parts show the inharmonic partials.}
    \label{fig:mel_partial}
\end{figure}


    \begin{figure}
        \centering
        \stepcounter{figure}
        \begin{minipage}[t]{0.12\textwidth}
            \centering
            \includegraphics[width=0.9\textwidth, trim= 90 32 520 30, clip]{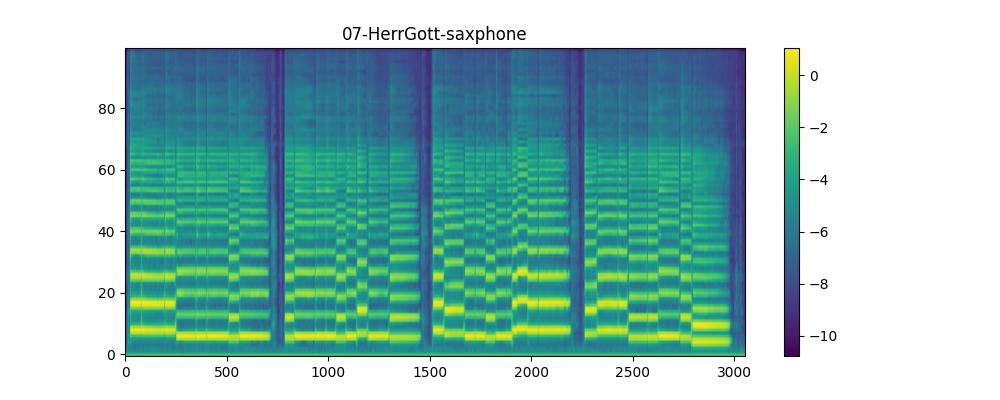}
            \Scaption{Source}
            \label{sfig:mel_cmp_origin}
        \end{minipage}%
        \begin{minipage}[t]{0.12\textwidth}
            \centering
            \includegraphics[width=0.9\textwidth, trim= 90 32 520 30, clip]{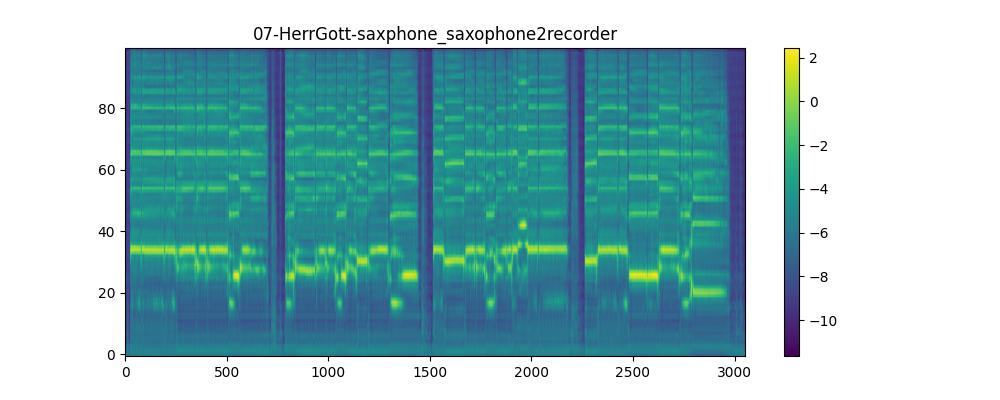}
            \Scaption{StarGAN}
            \label{sfig:mel_cmp_stargan}
        \end{minipage}%
        \begin{minipage}[t]{0.12\textwidth}
            \centering
            \includegraphics[width=0.9\textwidth, trim= 90 32 520 30, clip]{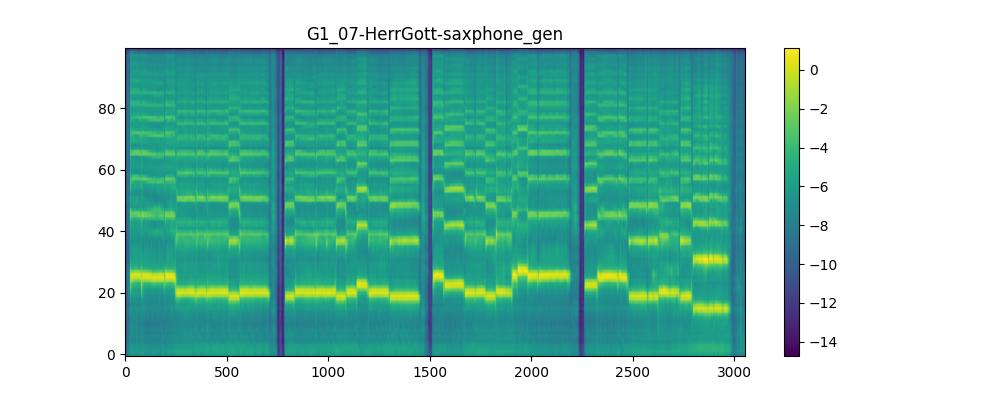}
            \Scaption{VAE-GAN}
            \label{sfig:mel_cmp_vaegan}
        \end{minipage}%
        \begin{minipage}[t]{0.12\textwidth}
            \centering
            \includegraphics[width=0.9\textwidth, trim= 125 44 722 48, clip]{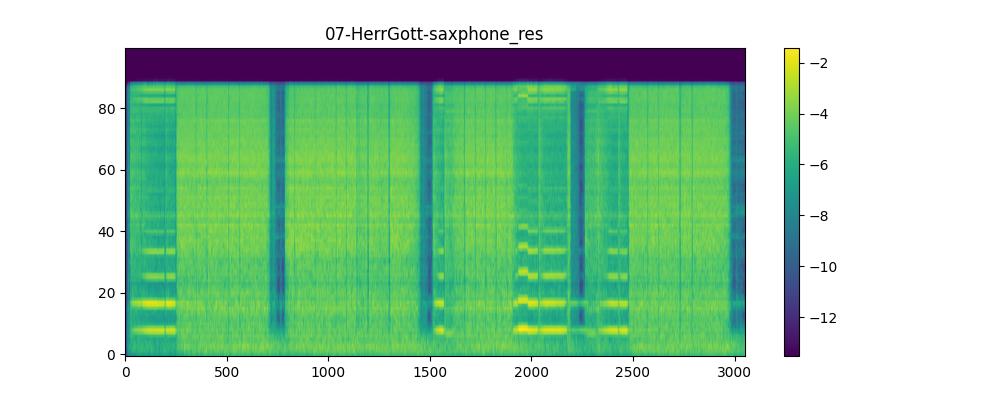}
            \Scaption{DDSP}
            \label{sfig:mel_cmp_ddsp}
        \end{minipage}
        \addtocounter{figure}{-1}
        \caption{The Mel spectrograms of a failed-music style transfer example. (a) The Mel spectrogram of the source audio, which is performed by a saxophone; (b) The Mel spectrogram of the converted audio (to failed recorder) by StarGAN; (c) The Mel spectrogram of the converted audio (to failed recorder) by VAE-GAN. (d) The Mel spectrogram of the converted audio (to failed recorder) by DDSP.}
        \label{fig:mel_cmp}
    \end{figure}

    \begin{table}[t]
    \begin{center}
    \begin{threeparttable}
        \caption{Wiener entropy of the URMP dataset and FR109 dataset. Vn., Cl., Sax., and Rec. represent violin, clarinet, saxophone, and recorder, respectively.}
        \label{tab:Wiener_entropy_dataset}
        \begin{tabular}{|c|c|c|}
            \hline
            Dataset & Instrument & Wiener entropy  \\
            \hline \hline
            URMP & Vn. / Cl. / Sax. & 0.0005 \\
            \hline
            FR109 & Rec. & 0.0345 \\
            \hline        
        \end{tabular}
    \end{threeparttable}
    \end{center}
    \end{table}

    \begin{table}[t]
    \begin{center}
    \begin{threeparttable}
        \caption{Wiener entropy of the style transfer results of StarGAN, VAE-GAN and DDSP on different target instruments. Vn., Cl., Sax., and Rec. represent violin, clarinet, saxophone, and recorder, respectively.}
        \label{tab:Wiener_entropy_models}
        \begin{tabular}{|c|c|c|c|c|}
            \hline
            Model/Target & Vn. & Cl. & Sax. & Rec. \\ \hline\hline
            StarGAN & 0.0007 & 0.0004 & 0.0002 & 0.0153 \\  \hline
            VAE-GAN & 0.0003 & 0.0003 & 0.0006 & 0.0159 \\ \hline
            DDSP & 0.0008 & 0.0008 & 0.0005 & 0.2041 \\ \hline
        \end{tabular}
    \end{threeparttable}
    \end{center}
    \end{table}


\section{Analysis}
    In this section, we analyze the results of failed recorder style transfer, and further compare the tasks between the conversion to well-played instruments (violin, clarinet, saxophone) and failed recorder.

\subsection{Mel spectrogram analysis}\label{subsec:mel_spec}
    To understand the challenge of failed-music style transfer, we first visualize the spectrograms of the failed recorder music in the FR109 dataset, as shown in \Cref{fig:mel_partial}. The red rectangular parts of the Mel spectrograms implies that the sound has \emph{inharmonic partials}, meaning that the frequencies of the overtones do not align with integer multiples of the fundamental frequency. This creates a more complex and less predictable timbre. These inharmonic partials are considered features of the failed recorder because they are present in the Mel spectrograms of every failed recorder sample. Such inharmonic partials are rarely found in well-played instrument performances.
    

    Next, we visualised an example of failed-music style transfer, Figure~\ref{fig:mel_cmp}\ref{sfig:mel_cmp_origin} shows the Mel spectrogram of source audio performed by a saxophone, in which there is no clear inharmonic partial. Figure~\ref{fig:mel_cmp}\ref{sfig:mel_cmp_stargan} and Figure~\ref{fig:mel_cmp}\ref{sfig:mel_cmp_vaegan} show the audio converted to a failed recorder by StarGAN and VAE-GAN, respectively. We can clearly see that inharmonic partials occur throughout the whole Mel spectrogram of StarGAN, showing that it does capture the characteristic of failed recorders and performs style transfer accordingly. For VAE-GAN, inharmonic partials can still be seen, but not as clearly as that of StarGAN. This shows that in this particular case, while both StarGAN and VAE-GAN do perform style transfer to some extent, StarGAN achieves a better style similarity to a failed recorder. Our informal listening test also confirms this observation.

    Based on \Cref{fig:mel_partial} and \Cref{fig:mel_cmp}, it can be seen that failed-music style transfer does show a clearly different characteristic to the style transfer of other well-played instruments. To achieve style transfer to failed music, a model has to generate audio with unique properties that do not usually occur in well-played music. Discussing such a task would help understand the performance and the limitation of a style transfer model in another aspect.

    

\subsection{Wiener entropy}
    Furthermore, we utilized the STFT-based \emph{Wiener entropy}~\cite{johnston1988transform} to quantify how much the noise-like sound is in the results produced by StarGAN and VAE-GAN, along with the Wiener entropy of the URMP dataset and FR109 dataset, which serve as the benchmark for real performance of well-played music and failed music. \Cref{tab:Wiener_entropy_dataset} shows the Wiener entropy of the URMP dataset and the FR109 dataset, i.e. well-played instrument music and failed recorder music, we can see that failed recorder music exhibits a higher proportion of noise-like characteristics compared to well-played instrument music. \Cref{tab:Wiener_entropy_models} shows the Wiener entropy of each of the models on different target instruments. We can see that when the target instruments are well-played instruments, the noise in the results from StarGAN and VAE-GAN are similar to the URMP dataset since their Wiener entropy is very similar. For failed recorder music, we can see that there is a gap between the Wiener entropy of FR109 and the Wiener entropy of StarGAN or VAE-GAN for converting to failed recorder, this shows that there is still room for improvement in converting music to the failed recorder style. 

\section{Conclusion and future work}\label{sec:conclusion}
    In this paper, we have conducted a series of experiments on the failed-music style transfer, and analysed the characteristics of this relatively special transfer in different aspects through various evaluations.
    
    Furthermore, we have released the FR109 dataset, consisting of failed recorder performances, which is useful for investigating the expressiveness of different style transfer model. Through this study, we hope to propose a music style transfer task that is different from the usual music style transfer task that pursues sound quality and accuracy, but rather a music style transfer task that is more versatile.
    

\printbibliography

\end{document}